\documentclass[twocolumn,floatfix,preprintnumbers,citeautoscript]{revtex4}
\usepackage{graphicx}
\usepackage{amsmath,amssymb,graphics}
\usepackage{bbm} 
\usepackage{accents} 
\usepackage{mathrsfs}	
\usepackage[T1]{fontenc}
\usepackage{psfrag}
\usepackage{upgreek}
\usepackage{verbatim}

\newcommand{\slashed}[1]{\displaystyle{\not}{#1}}

\begin{document} \setlength{\unitlength}{1in}

\preprint{TTK-12-08, CAS-KITPC/ITP-315}
 
\title{\Large{Sterile Neutrinos as the Origin of Dark and Baryonic Matter}}

\author{{\bf\normalsize Laurent Canetti,$^1$
Marco Drewes,$^{2,3}$
Mikhail Shaposhnikov,$^1$}\\[0.5cm]
{\it\normalsize
$^1$Institute de Th\'eorie des Ph\'enom\`enes Physiques
EPFL, CH-1015 Lausanne, Switzerland}\\
{\it\normalsize
$^2$Physik Department T70, Technische Universit\"at M\"unchen,} {\normalsize \it James Franck Stra\ss e 1, D-85748 Garching, Germany\\ 
$^3$Institut f\"ur theoretische Teilchenphysik und Kosmologie, RWTH Aachen, D-52056 Aachen, Germany
}\\
[0.15cm]
}
 \begin{abstract}  
We demonstrate for the first time that three sterile neutrinos alone can simultaneously explain neutrino oscillations, the observed dark matter and the baryon asymmetry of the Universe without new physics above the Fermi scale. The key new point of our analysis is leptogenesis after sphaleron freeze-out, which leads to resonant dark matter production, evading thus the constraints on sterile neutrino dark matter from structure formation and x-ray searches. We identify the range of sterile neutrino properties that is consistent with all known constraints. We find a domain of parameters where the new particles can be found with present day experimental techniques, using upgrades to  existing experimental facilities.
\end{abstract}
\pacs{\dots}

\maketitle

{\it Introduction - } The standard model (SM) of particle physics and theory of general relativity describe correctly almost all phenomena observed in nature. Only a handful of experimental facts definitely involve particle physics beyond the SM: neutrino oscillations (NOs), dark matter (DM), and the baryon asymmetry of the Universe (BAU), which is responsible for today's remnant baryon density $\Omega_{\rm B}$. In Refs. \cite{Asaka:2005an,Asaka:2005pn}, it has been suggested that all of them may be explained when the matter content of the SM is complemented by three right-handed neutrinos with masses below the electroweak scale. Different authors have investigated aspects of this idea
\cite{Shaposhnikov:2008pf,
Laine:2008pg,
Bezrukov:2008ut,
Boyarsky:2009ix,
Roy:2010xq,
Asaka:2005pn,
Asaka:2005an,
Boyarsky:2005us,
Asaka:2006rw,
Bezrukov:2005mx,
Shaposhnikov:2006nn,
Asaka:2006ek,
Boyarsky:2006zi,
Boyarsky:2006fg,
Boyarsky:2006jm,
Shaposhnikov:2006xi,
Gorbunov:2007ak,
Gorbunov:2007zz,
Bezrukov:2007qz,
Anisimov:2008qs,
Canetti:2010aw,
Asaka:2010kk,
Asaka:2006nq,
Gorkavenko:2009vd,
Asaka:2011pb,
Ruchayskiy:2011aa,
Gorkavenko:2012mj,
Ruchayskiy:2012si,
Drewes:2012ma,
Merle:2012xq}. However, to date, it has not been verified that all requirements can be fulfilled {\it simultaneously}. Claims made in Ref. \cite{Asaka:2005pn} were based on estimates and turn out to be premature from today's point of view due to constraints on the properties of DM sterile neutrinos from Ly$_\alpha$ and x-ray observations that were not known at that time. These constraints can be resolved if DM production is enhanced by a lepton asymmetry generated after sphaleron freeze-out. We performed the first quantitative study of this process to identify the range of sterile neutrino parameters that allow us to explain at once NO, DM, and the BAU. In this letter, we mainly present results; details are given in a more detailed publication \cite{Canetti:2012kh}. The centrepiece of our analysis is the study of neutrino abundances in the early Universe from hot big bang initial conditions to temperatures $\sim 50$ MeV. We combine the results with bounds from direct searches for sterile neutrinos and constraints from big bang nucleosynthesis (BBN), which we reexamined in the face of recent data from neutrino experiments. We verify for the first time that right-handed neutrinos with {\em experimentally accessible masses and mixings} can solve all these outstanding problems without any new physics above the Fermi scale. We identify the experimentally interesting parameter region for future searches. 

{\it The Neutrino Minimal Standard Model ($\nu$MSM) -} The scenario outlined above is realized within the $\nu$MSM, described by the Lagrangian
\begin{eqnarray}
	\label{nuMSM_lagrangian}
	\mathcal{L}_{\nu MSM} =\mathcal{L}_{SM}+ 
	i \bar{\nu}_{R}\slashed{\partial}\nu_{R}-
	\bar{L}_{L}F\nu_{R}\tilde{\Phi} -
	\bar{\nu}_{R}F^{\dagger}L_L\tilde{\Phi}^{\dagger} 
	\nonumber\\ -{\rm \frac{1}{2}}(\bar{\nu_R^c}M_{M}\nu_{R} 
	+\bar{\nu}_{R}M_{M}^{\dagger}\nu^c_{R}). 
\end{eqnarray}
We have suppressed flavor and isospin indices. $\mathcal{L}_{SM}$ is the Lagrangian of the SM. $F$ is a matrix of Yukawa couplings, and $M_{M}$ is a Majorana mass term for the right-handed neutrinos $\nu_{R}$. $L_{L}=(\nu_{L},e_{L})^{T}$ are the left-handed lepton doublets in the SM, and $\Phi$ is the Higgs doublet. We chose a basis where the charged lepton Yukawa couplings and $M_{M}$ are diagonal. The Lagrangian (\ref{nuMSM_lagrangian}) is well-known in the context of the seesaw mechanism \cite{Minkowski:1977sc,Yanagida:1980xy,GellMann:1980vs,Mohapatra:1979ia}. In the $\nu$MSM, the eigenvalues of $M_M$ are below the electroweak scale \footnote{For a summary of possible origins of a low seesaw scale see \cite{Abazajian:2012ys} and references therein, see also \cite{Iso:2012jn,Harigaya:2012bw,Nemevsek:2012cd,Araki:2011zg,Adulpravitchai:2011rq,Merle:2011yv,Barry:2011wb,Lindner:2010wr,McDonald:2010jm,Kusenko:2010ik,Bezrukov:2009th,Kadota:2007mv,Chen:2006hn}.}. This mass pattern is required to simultaneously explain BAU and DM; at the same time, it avoids the ``hierarchy problem'' of the SM in the scale-invariant version of the $\nu$MSM \cite{Shaposhnikov:2008xb,Shaposhnikov:2008xi}. The $\nu$MSM is motivated by the principle of minimality; in comparison with the SM, there is no modification of the gauge group, the number of fermion families remains unchanged, and no new energy scale {\em above} the Fermi scale is introduced \cite{Shaposhnikov:2007nj,Boyarsky:2009ix}.

In the $\nu$MSM, neutrino masses are generated from Dirac masses $m_D=Fv$ and Majorana masses $M_M$ by the seesaw mechanism ($v$ is the Higgs vacuum expectation value). In the limit $M_M\gg m_D$, there are two distinct sets of neutrino mass eigenstates. The block diagonalization of the full mass matrix yields mass matrices for active and sterile neutrinos $m_\nu=-\theta M_M \theta^T$ and  $M_N=M_M + \frac{1}{2} (\theta^{\dagger} \theta M_M + M_M^T \theta^T \theta^{*} )$, respectively. The active mass eigenstates $\upnu_i$ with masses $m_i$ are mainly mixings of the SM neutrinos $\nu_{L,\alpha}$; the remaining three sterile neutrinos $N_I$ with masses $M_I$ are mainly mixings of $\nu_{R,I}$. Transitions between both are suppressed by the active-sterile mixing matrix  $\theta=m_D M_M^{-1}$. 

In the following, we distinguish between three different scenarios. In scenario I, no physics beyond the $\nu$MSM is required to explain simultaneously NO, $\Omega_{\rm DM}$ and $\Omega_{\rm B}$, as outlined in the Introduction. One sterile neutrino ($N_1$) constitutes all DM. This implies that (a) its mass and mixing are consistent with astrophysical constraints and (b) thermal production can account for the observed $\Omega_{\rm DM}$. The other two ($N_{2,3}$) produce the BAU and generate active neutrino masses via the seesaw mechanism. In scenario II, the roles of the $N_I$ are the same, but we drop requirement (b), i.e., we assume that DM is made of $N_1$, that was produced by some unspecified mechanism. In scenario III, we consider the $\nu$MSM as a theory of baryogenesis only and drop any constraints related to DM.

To explain the observed DM density $\Omega_{\rm DM}$ in scenarios I and II, the lifetime of $N_1$ must be larger than the age of the Universe. Its decay leaves a distinct x-ray line of energy $M_1/2$  that can be searched for \cite{Dolgov:2000ew,Abazajian:2001nj} with x-ray satellites. Combining x-ray observations with simulations of structure formation and phase space arguments, it was found that $N_1$ mass and mixing are constrained to $1$ keV$<M_1\lesssim 50$ keV and $10^{-13}\lesssim\sin^2(2\theta_{\alpha 1})\lesssim 10^{-7}$ \cite{Boyarsky:2009ix}. The seesaw relation $m_i\sim -\theta M_M \theta^T$ implies that the coupling of $N_1$ is too small to contribute significantly to the active neutrino mass matrix, and one active neutrino is effectively massless in the $\nu$MSM \cite{Asaka:2005an,Boyarsky:2006jm}. 

\begin{figure} 
  \centering
    \includegraphics[width=8cm]{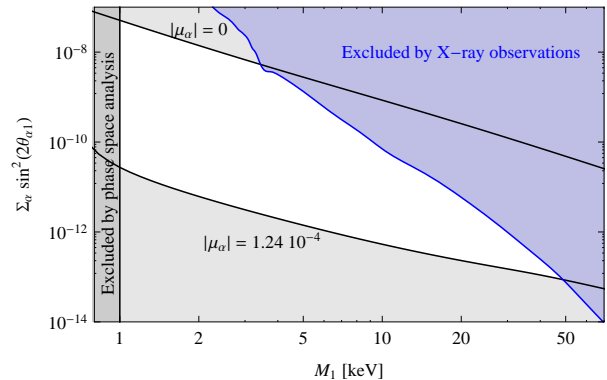}
    \caption{Different constraints on $N_1$ mass and mixing in scenario I. The blue region is excluded by x-ray observations, the dark gray region $M_1< 1$ keV by the Tremaine-Gunn bound \cite{Tremaine:1979we,Boyarsky:2008ju,Gorbunov:2008ka}. For points on the upper solid black line, the observed $\Omega_{\rm DM}$ is produced for $\mu_{\alpha}=0$ \cite{Laine:2008pg}; points on the lower solid black line give the correct $\Omega_{\rm DM}$ for $|\mu_{\alpha}|=1.24\cdot10^{-4}$, the maximal asymmetry we found at $T=100$ MeV. The region between these lines is accessible for $0\leq |\mu_\alpha|\leq 1.24\cdot10^{-4}$. Observations of the matter distribution in the Universe constrain the DM free streaming length. Without resonant production ($\mu_\alpha=0$), this implies that $M_1>8$ keV \cite{Boyarsky:2008mt}, which excludes the upper black curve and makes resonant production necessary.  Combining both production mechanisms ($|\mu_\alpha|\gtrsim10^{-5}$), this bound relaxes to $M_1>2$ keV \cite{Boyarsky:2008mt}. However, we do not display it here because it depends on $\mu_\alpha$ in a complicated way and the calculation currently includes considerable uncertainties \cite{Boyarsky:2008mt}, cf. also \cite{Boyanovsky:2008nc,Boyanovsky:2010pw}. 
\label{DMexclusion}}
\end{figure}

In scenario I, $N_1$ must be produced thermally in the early Universe due to active-sterile mixing \cite{Dodelson:1993je}. In the absence of lepton asymmetries, $\mu_\alpha=0$, the resulting spectrum was found in \cite{Asaka:2006nq}. For $\mu_\alpha\neq0$, the $N_1$ dispersion relation in the primordial plasma is modified, which results in a resonantly amplified $N_1$ production \cite{Shi:1998km,Laine:2008pg,Shaposhnikov:2008pf,Wu:2009yr}. This adds a nonthermal, colder component to the $N_1$ momentum distribution. x-ray observations, structure formation simulations, and Ly$_\alpha$ forest data \cite{Boyarsky:2008mt,Boyarsky:2009ix,Boyarsky:2008xj,Seljak:2006qw} suggest that if $\mu_\alpha=0$, $N_1$ cannot account for the observed DM (see figure \ref{DMexclusion}). Then the presence of considerable lepton asymmetries $|\mu_\alpha|\gtrsim 8\cdot 10^{-6}$ \cite{Laine:2008pg} becomes a necessary condition for sterile neutrino DM production. 

The thermal history of the universe in the $\nu$MSM differs from that of the SM, as new interactions generate lepton asymmetries $\mu_\alpha\neq0$ during  production, oscillations, freeze-out, and decay of $N_I$, when all Sakharov conditions \cite{Sakharov:1967dj} are fulfilled. No significant lepton asymmetries or $N_I$ abundances are created during reheating after inflation due to the smallness of $F$ \cite{Bezrukov:2008ut}. Baryogenesis occurs via sterile neutrino oscillations during their thermal production \cite{Akhmedov:1998qx,Asaka:2005pn} in processes as $t\bar{t}\rightarrow\upnu N$ at $T\gtrsim T_{\rm sph}$, where the temperature of sphaleron freeze-out is $T_{\rm sph} \sim 140$ GeV for a Higgs mass $m_H=126$ GeV. Although the total lepton number violation is suppressed by $M_I/T\ll1$, opposite sign asymmetries are created in the sterile and active flavors. The latter are partly converted into a BAU by sphaleron processes \cite{Kuzmin:1985mm}. To explain the observed BAU \cite{Canetti:2012zc}, a lepton asymmetry $\mu_\alpha\sim 10^{-10}$ is required at the sphaleron freeze-out ($T\sim T_{\rm sph}$). In scenarios I and II, only $N_{2,3}$ are produced in significant amounts at $T\gtrsim T_{\rm sph}$, as the $N_1$ coupling is constrained to be tiny. Soon after, they reach equilibrium and $\mu_\alpha$ are washed out. They exit equilibrium when $l\bar{l}\rightarrow\upnu N$ scatterings freeze out ($T\sim$ few GeV) and decay subsequently ($T\lesssim 1$ GeV). These nonequilibrium processes create new lepton asymmetries at a late time. The DM production in scenario I is amplified resonantly at temperatures around $T\sim 100$ MeV due to the presence of these asymmetries.

We first focus on scenario I. The two requirements (i) $\mu_\alpha \sim 10^{-10}$ at $T\sim T_{sph}$ (for BAU) and (ii) $|\mu_\alpha|\gtrsim 8\cdot 10^{-6}$ at $T\sim 100$ MeV (for DM) can be used to constrain the properties of sterile neutrinos. Since $N_1$ does not contribute to high and low temperature leptogenesis, this can be done in an effective theory with two sterile neutrinos, $N_{2,3}$. This theory contains $11$ new parameters in addition to the SM. They can be chosen as two active neutrino masses $m_i$ and three mixing angles, a Dirac and a Majorana phase, two Majorana masses in $M_M$ and one extra complex parameter, associated with CP-violation in the sterile sector.  

Condition (ii) implies much stronger constraints than (i), so we only consider it in what follows. For the allowed Yukawa couplings, the asymmetry (ii) can only be generated if the CP-violating terms are resonantly amplified by a mass degeneracy between $M_{2,3}\simeq M$ \cite{Shaposhnikov:2008pf}. The asymmetry generation is most efficient for $\Gamma_N\sim {\rm H}\sim \omega$, where $\Gamma_N$ is the thermal $N_I$ width, H is the Hubble rate, and $\omega$ is the frequency of $N_{2,3}$ oscillations. It is related to the {\em physical} mass splitting $\delta M$ at the time of low-temperature lepton asymmetry generation via $\omega\sim M\delta M/T$ if $M\lesssim T$ or simply $\omega=\delta M$ if $T\lesssim M$. Due to the interplay of thermal, Dirac and Majorana masses, $\delta M$ is a complicated function of $M_M={\rm diag}(M-\Delta M/2,M+\Delta M/2)$, $F$ and $T$ \cite{Canetti:2012kh}. The required $\delta M$ should be smaller than $10^{-6} {\rm eV}$ for $M=2$ GeV. Since this is much smaller than active neutrino masses, {\em two} unknown parameters of the $\nu$MSM are almost fixed by this constraint. For $M\sim 2$ GeV, the splitting of the Majorana masses $\Delta M$ must be equal with one part in $10^4-10^6$ to the mass difference of the active neutrinos \cite{Roy:2010xq}. In addition, the combination $|Re (m^\dagger_D m_D)_{23}|/M$ must be smaller than active neutrino masses by $\sim 4-6$ orders of magnitude \cite{Canetti:2012kh}. These tunings ensure that the Higgs induced contribution to $\delta M$ cancels the Majorana term $\Delta M$. Scenario I can only be realized within the {\it constrained} $\nu$MSM defined by these tunings. In the constrained $\nu$MSM, 7 out of 11 parameters are almost fixed either by experimental data or by the requirement (ii). The remaining four parameters are the common mass $M$ of $N_{2,3}$, two CP-violating phases in the active neutrino mass matrix, and yet one extra CP-violating parameter in the sterile neutrino sector.

The high degree of tuning $\delta M/M\lesssim 10^{-13}$ in scenario I is not understood theoretically. Some speculations can be found in \cite{Shaposhnikov:2006nn,Shaposhnikov:2008pf,Roy:2010xq,He:2009pt}. However, the origin of this fine tuning plays no role for the present work. Scenario II only requires the weaker condition (i) that can be achieved by a much weaker tuning in $\Delta M/M\sim 10^{-3}$ \cite{Canetti:2010aw}. In scenario III, all three sterile neutrinos can participate in baryogenesis. Due to the additional sources of CP violation, there is no need for a mass degeneracy \cite{Drewes:2012ma}. Note that this also implies that no degeneracy is needed in scenario II if more than three fields $\nu_{R,I}$ are added to the SM.

{\it Method and Results. -} The rates of interaction of the SM fields exceed those of sterile neutrinos $N_I$ and the rate of the Universe expansion. Therefore, the SM sector can be described by four numbers: the temperature $T$ and the asymmetries $\mu_\alpha$. The effect of the $N_I$ on the time evolution of temperature and on the effective number of degrees of freedom is negligible. The state of $N_I$ can be described by {\it matrices of density} $\rho_N$ and $\rho_{\bar{N}}$, commonly used in neutrino physics \cite{Sigl:1992fn}, which allow us to incorporate coherences and oscillations between the two flavors and are probably sufficient for our purpose, cf.
\cite{Sigl:1992fn,Anisimov:2010aq,Anisimov:2010dk,
Beneke:2010dz,Beneke:2010wd,Garny:2011hg,Buchmuller:2000nd,Anisimov:2008dz,
Garny:2009rv,Garny:2010nz,Garbrecht:2011aw,Drewes:2012ma,Garny:2009qn,
Drewes:2010pf,Drewes:2012qw,Frossard:2012pc,Asaka:2011wq,Boyanovsky:2007as,
Boyanovsky:2007zz,Boyanovsky:2006it,Bezrukov:2012as}. The diagonal elements of the $2\times2$ matrices $\rho_N$ and $\rho_{\bar{N}}$ are the abundances of particles and antiparticles, respectively, defined as the helicity states of the Majorana fields $N_I$. The time evolution of neutrino abundances is governed by the kinetic equations
\begin{eqnarray}
\label{kinequ1}
i \frac{d\rho_{N}}{d T}&=&[H, \rho_{N}]-\frac{i}{2}\{\Gamma_N, \rho_{N} 
- \rho^{eq}\} +\frac{i}{2} \mu_\alpha{\tilde\Gamma^\alpha_N}~,\\
i \frac{d\rho_{\bar{N}}}{d T}&=& [H^*, \rho_{\bar{N}}]-\frac{i}{2}
\{\Gamma^*_N, \rho_{\bar{N}} - \rho^{eq}\} -
\frac{i}{2} \mu_\alpha{\tilde\Gamma_N^{\alpha *}}~,\label{kinequ2}\\
i \frac{d\mu_\alpha}{d T}&=&-i\Gamma^\alpha_L \mu_\alpha +
i {\rm tr}\left[{\tilde \Gamma^\alpha_L}(\rho_{N} -\rho^{eq})\right]
\nonumber\\ 
&&-i {\rm tr}\left[{\tilde \Gamma^{\alpha*}_L}(\rho_{\bar{N}} 
 -\rho^{eq})\right]~.
\label{kinequ3}
\end{eqnarray}
Here, $\rho^{eq}$ is the equilibrium density matrix, $H$ is the dispersive part of the finite temperature effective $N_I$ Hamiltonian, and $\Gamma_N$, $\Gamma^\alpha_L$ and ${\tilde \Gamma^\alpha}_L$ are rates that are responsible for dissipative effects, which are calculated from thermal field theory \cite{Canetti:2012kh}. They describe sterile neutrino production, oscillations, freeze-out and decay. Due to the various involved time scales, the dependence of the asymmetries on model parameters can only be estimated under certain assumptions \cite{Shaposhnikov:2008pf,Beneke:2010dz,Garny:2011hg,Drewes:2012ma} that are too simplifying for a quantitative study.

We focussed on scenarios I and II. The most important properties of $N_{2,3}$ from an experimental viewpoint are their masses $M_{2,3}\simeq M$ and mixings with active neutrinos. The latter can be parameterized by the quantity $U^2\equiv{\rm tr}(\theta^\dagger\theta)$. In order to identify the range of $M$ and $U^2$ consistent with conditions (i) and (ii), we calculated the lepton asymmetries from $T\gg T_{sph}$ down to $T=100$ MeV as a function of unknown parameters identified above and varied the others within admitted $1 \sigma$ uncertainties. 

Fixing all known neutrino parameters to the values given in \cite{Fogli:2011qn}\footnote{Using the more recent values for $\uptheta_{13}$ found in \cite{An:2012eh,Ahn:2012nd} has no visible effect on our results.}, we first identified the Dirac and Majorana phases that maximize the produced asymmetries at $T=T_{sph}$ and $T=100$ MeV in different regions in the parameter space. We then scanned for all possible values of the remaining parameters. We performed the analysis several times with different grids. This allows us to identify the parameter regions where condition (i) or (ii) or both can be fulfilled. They correspond to sterile neutrino properties for which the $\nu$MSM can, along with NOs, explain the observed $\Omega_{\rm B}$, $\Omega_{\rm DM}$ or both. We studied the mass range $1$ MeV $\leq M\leq 10$ GeV; for bigger masses it is very unlikely that $N_{2,3}$ can be found experimentally in the near future.

The $\nu$MSM parameter space is also constrained by direct searches
for sterile neutrinos \cite{Gorbunov:2007ak,Asaka:2012bb,Lello:2012gi,
Yamazaki:1984sj,Daum:2000ac,Bernardi:1985ny,Bernardi:1987ek,Vaitaitis:1999wq,
Astier:2001ck,Atre:2009rg,Ruchayskiy:2011aa,Ando:2010ye,Li:2010vy,Liao:2010yx,
Kersten:2007vk}. Here, we focus on the most relevant bounds, coming from the NuTeV \cite{Vaitaitis:1999wq}, CHARM \cite{Bergsma:1985is} and CERN PS191 \cite{Bernardi:1985ny,Bernardi:1987ek} experiments. The experimental constraints on active-sterile mixing have recently been interpreted in the context of the seesaw Lagrangian (\ref{nuMSM_lagrangian}) \cite{Asaka:2011pb,Ruchayskiy:2011aa}, cf. also  \cite{Lello:2012gi,Asaka:2012bb}. We used bounds on $\theta$ imposed by the negative results of \cite{Bernardi:1985ny,Vaitaitis:1999wq,Bergsma:1985is,Britton:1992xv,Hayano:1982wu,Bryman:1996xd,Abela:1981nf,Daum:1987bg}, provided by the authors of \cite{Ruchayskiy:2011aa}, as input and numerically scanned the space of unknown model parameters to identify all combinations of $M$ and $U^2$ compatible with experiment.

Finally, it is a necessary requirement that $N_{2,3}$ have decayed sufficiently long before BBN that their decay products do not affect the abundances of light elements. We estimate the inverse $N_{2,3}$ lifetime $\tau$ by as $\tau^{-1}\simeq\frac{1}{2}{\rm tr}\Gamma_N$ at $T=1$ MeV. This is justified as $N_{2,3}$ oscillate rapidly around the time of BBN. We varied all free parameters to identify the region in the $M$-$U^2$ plane consistent with the condition $\tau<0.1$s. 
\begin{figure}
  \centering
    \includegraphics[width=8cm]{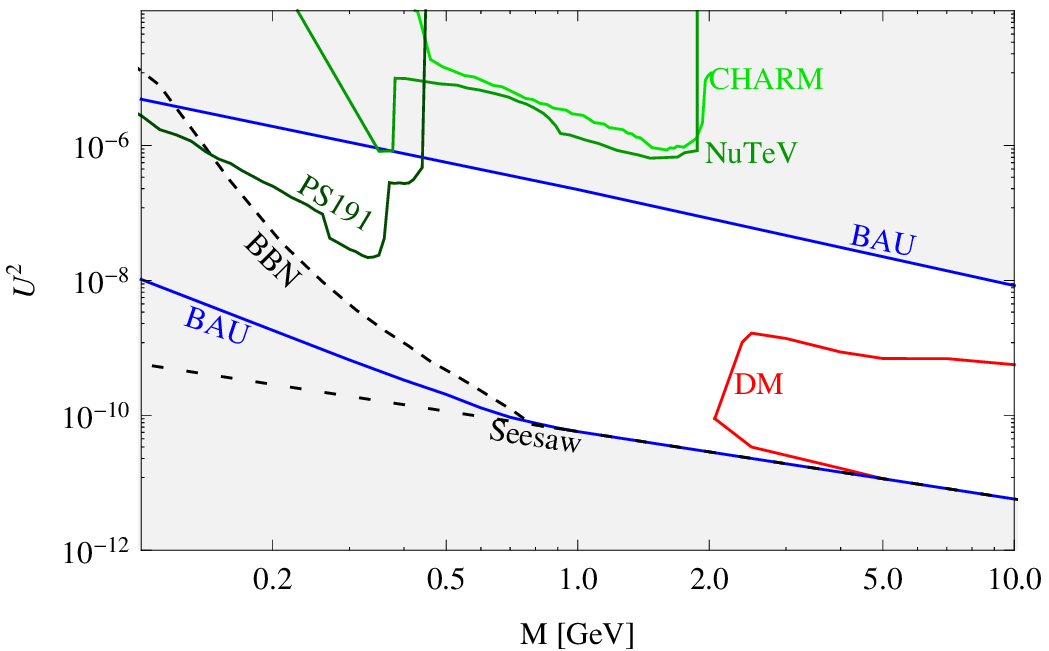}\\
    \includegraphics[width=8cm]{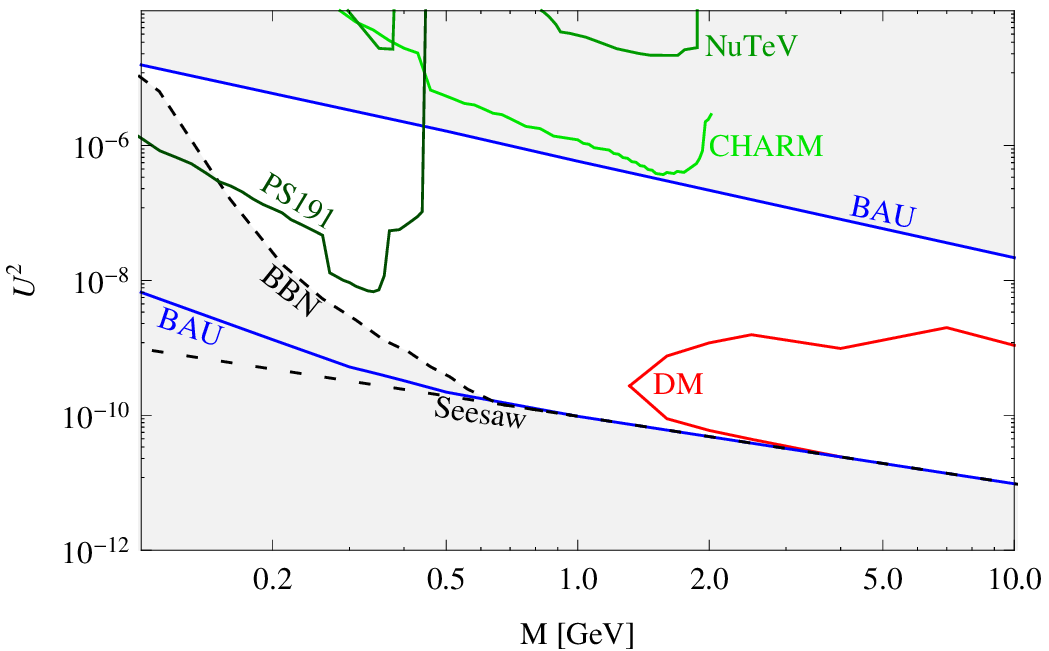}
    \caption{Constraints on $N_{2,3}$ masses $M_{2,3}\simeq M$ 
    and mixing $U^2\equiv{\rm tr}(\theta^\dagger\theta)$ 
    in scenarios I (red line) and II (blue line)
for normal (upper panel) and inverted (lower panel) hierarchy 
    of neutrino masses. In the regions within the blue and red 
    lines, no physics beyond the $\nu$MSM is needed to explain 
    the observed $\Omega_{\rm B}$ and $\Omega_{\rm DM}$, respectively.
    \label{BA_DM_theta}}
\end{figure} 

Our results are summarized in Figs. \ref{DMexclusion} and \ref{BA_DM_theta}.  Figure \ref{BA_DM_theta} shows constraints on $N_{2,3}$ mass and mixing coming from experiments (green lines), BBN (dashed black line), neutrino oscillations experiments (black dashed line that is labeled seesaw), and cosmology. Scenario II can be realized in the region between the blue lines that are labeled BAU. The difference to the result found in \cite{Canetti:2010aw} is mainly due to $\uptheta_{13}\neq0$. The region within the red line allows us to produce the observed $\Omega_{\rm DM}$. It has been determined for the first time in this work. Although the values for the CP-violating phases and $\Delta M$ that maximize the efficiency of baryogenesis and DM production are very different, the region in which $\Omega_{\rm B}$ and $\Omega_{\rm DM}$ can be explained simultaneously (scenario I) almost coincides with the area inside the red line.   In most of the parameter space, the relevant CP-violation comes mainly from the sterile sector and not from the phases in the Pontecorvo-Maki-Nakagawa-Sakata matrix. The masses of $N_{2,3}$ are correspondingly bounded from below by $1.3$ and $2.0$ GeV for the inverted and normal hierarchies. For the lower mass range, sterile neutrinos can be created in decays of beauty and charmed mesons, which is crucial for experimental searches \cite{Gorbunov:2007ak}. The two larger eigenvalues of $F^\dagger F$ can vary between $\sim 10^{-12}$ and $\sim 10^{-17}$ to simultaneously explain $\Omega_{\rm B}$ and $\Omega_{\rm DM}$. Typical values correspond to Yukawa couplings of $N_{2,3}$ of the order $10^{-6}-10^{-7}$, smaller than the electron Yukawa by 1-2 orders of magnitude. The $N_1$ Yukawa couplings are required to be of the order  $10^{-11}-10^{-12}$, smaller than those of  $N_{2,3}$ by 5 orders of magnitude, which is comparable to the ratio between down and top quark Yukawa couplings. 

Solving Eqs. (\ref{kinequ1})-(\ref{kinequ3}) allows us to determine the maximal lepton asymmetry generated in the $\nu$MSM. Its value imposes a lower bound on the mixing of the DM candidate $N_1$ in scenario I. Our result for $M<10$ GeV, shown in Fig. \ref{DMexclusion} along with astrophysical bounds, is about $1$ order of magnitude larger than previous estimates \cite{Shaposhnikov:2008pf}. This considerably eases the ultimate goal of x-ray searches for $N_1$.

{\it Conclusion - } We performed the first complete systematic study of the $\nu$MSM parameter space, bringing together cosmological, astrophysical and experimental constraints. Our results can be summarized as follows: (1) Right-handed neutrinos alone can be the common origin of neutrino oscillations, DM and the BAU; (2) for a range of model parameters, these particles can be found using present day experimental and observational techniques. 

The DM candidate $N_1$ can be searched for astrophysically, using high resolution x-ray spectrometers to look for the emission line from its decay in DM dense regions. The seesaw partners $N_{2,3}$ can either be discovered as missing energy in the decay of mesons or by creating them in a beam-dump experiment and looking for their decay in a nearby detector \cite{Gorbunov:2007ak}. Depending on the mass $M$, different facilities could be used or upgraded for this search, including the CERN SPS beam and NA62 experiment, LHCb, MINOS, J-PARC or LBNE at FNAL. The necessary experiments are challenging due to strong constraints on the mixing angle $U^2$ coming from cosmology.
\\ 
\newline
{\bf Acknowledgements} - We are grateful to M.~Laine for numerical
data on finite temperature scattering rates as well as O.~Ruchayskiy
and  A.~Ivashko for sharing their expertise on experimental bounds. 
We also thank O.~Ruchayskiy, T.~Asaka, D.~Gorbunov and A.~Boyarsky for
their comments. We are grateful to T.~Frossard for collaboration at
initial stage of this work. This work was supported by the Swiss
National Science Foundation, the Gottfried Wilhelm Leibniz program of
the Deutsche Forschungsgemeinschaft  and the Project of Knowledge
Innovation Program  of the Chinese Academy of Sciences grant
KJCX2.YW.W10.

\bibliographystyle{apsrev}
\bibliography{all}

\end{document}